# Variability Effects in Graphene: Challenges and Opportunities for Device Engineering and Applications

Guangyu Xu, *Member, IEEE,* Yuegang Zhang, *Member, IEEE*, Xiangfeng Duan, Alexander A. Balandin, *Fellow, IEEE* and Kang L. Wang, *Fellow, IEEE*

*Abstract*— Variability effects in graphene can result from the surrounding environment and the graphene material itself, which form a critical issue in examining the feasibility of graphene devices for large-scale production. From the reliability and yield perspective, these variabilities cause fluctuations in the device performance, which should be minimized via device engineering. From the metrology perspective, however, the variability effects can function as novel probing mechanisms, in which the 'signal fluctuations' can be useful for potential sensing applications. This paper presents an overview of the variability effects in graphene, with emphasis on their challenges and opportunities for device engineering and applications. The discussion can extend to other thin-film, nanowire and nanotube devices with similar variability issues, forming general interest in evaluating the promise of emerging technologies.

*Index Terms*—graphene; variability effects; device engineering; sensing applications; interface traps; low frequency noise; edge disorder; device scalability; metrology

## I. INTRODUCTION

GRAPHENE draws considerable interest in electronics, photonics and multiple cross-fields owing to its combination of exceptional properties [1] - high carrier mobility [2, 3], atomically-thin planar structure [4], linear dispersion of Dirac fermions [5], high mechanical strength[6],

Manuscript received May 3, 2012. This work was supported in part by SRC-DARPA MARCO Focus Center on Functional Engineered Nano Architectonics (FENA) and the U.S. Department of Energy under Contract No. DE-AC02-05CH11231.

Guangyu Xu was with the Department of Electrical Engineering, University of California at Los Angeles, Los Angeles, CA 90095 USA. He is now with School of Engineering and Applied Sciences, Harvard University, Cambridge, MA 02138 USA (email: guangyu@seas.harvard.edu).
Yuegang Zhang was with the Molecular Foundry, Lawrence Berkeley National Laboratory, Berkeley, CA 94720 USA. He is now with Suzhou Institute of Nano-Tech and Nano-Bionics, Chinese Academy of Sciences, Suzhou 215123 China (email:yzhang5@lbl.gov, ygzhang2012@sinano.ac.cn).
Xiangfeng Duan is with the Department of Chemistry and Biochemistry, University of California at Los Angeles, Los Angeles, CA 90095 USA (email: xduan@chem.ucla.edu).
Alexander. A. Balandin is with the Department of Electrical Engineering and Materials Science and Engineering Program, University of California at Riverside, Riverside, CA 92521 USA (email: balandin@ee.ucr.edu).
Kang L. Wang is with the Department of Electrical Engineering, University of California at Los Angeles, Los Angeles, CA 90095 USA (email: wang@ee.ucla.edu).

high thermal conductivity [7] and potential low-cost [8-10], among others [11]. Raised by the rapid progress of material synthesis and fabrication techniques [9, 12-15], graphene has shown its potential in wafer-scale radio-frequency analog circuits [16-23], broadband photodetection [24-26], electronic circuit interconnects [27-29], thermal management [30-33] and subnanometre trans-electrode membrane for DNA detection [34-36]. To date, graphene-based materials, generally referred to as graphene, graphene nanoribbon, graphene oxide and some others[1], are the focus of nanoscience and nanotechnology societies, with continuous efforts on exploring their versatile applications [1, 11, 37-40].

Increasing process variability poses a major challenge to the continued scaling of semiconductor technology (e.g. limits the reliability and yield); addressing this issue requires optimization of both device and circuit designs [41-47]. Variability sources in the standard complementary metal-oxide-semiconductor (CMOS) process can be categorized according to their spatial characteristics, time-scales, physical/environmental origins and systematic/random components [42, 43, 46]. And the nature of variability is likely to change with the progress of innovative materials, fabrication methods and device structures in the targeted applications [46, 48-52]: some variability sources may diminish while others may emerge; some can be minimized via device engineering (e.g. variation in nominal lengths/widths) while others are limited by the material imperfection (e.g. interface roughness and dopant fluctuation). The identification, characterization and evaluation of variability effects in emerging technologies are essential in examining their ultimate promises for large-scale production [53-56].

Motivated by the potential of graphene as a material candidate to incorporate into silicon devices for high-speed electronics and integrated photonics [40], there has been considerable interest on the variability effects in graphene and graphene-related technology [12, 51, 57-60]. At the moment, efforts are made on the characterization of graphene variabilities in the prototyped device structures and the evaluation of their effects on the device performance [60-65]. However, a systematic discussion of the variability sources in graphene and their impact on circuits and systems are rare. From the reliability perspective, these variabilities in graphene cause device fluctuations and are detrimental to the yield in large-scale production. It is therefore critical to seek ways of minimizing their effects via device engineering (e.g. adjusting the process flow or device structure), one topic that is being heavily investigated [12, 60, 63, 64, 66-70]. From the metrology perspective, on the other hand, we demonstrate that variabilities in graphene can function as novel probing mechanisms [58, 65, 71-75], where the 'signal fluctuations' can be useful for potential sensing applications. This role of graphene variabilities is of both fundamental and practical interest [76-78], extendable to other thin-films or nanomaterials [79-82], and may be employed in developing novel metrology applications.

With this introduction, here we review the status and prospects of variability effects in graphene, discussing their challenges and opportunities for device engineering and applications, respectively. Section II provides an overview of variability sources in graphene, with emphasis on their concepts, categorizations and the comparison with those in silicon devices. Two broad classes of variabilities in graphene, those from the environmental disturbance and those from the material imperfection, are described with typical examples. For the first class, Section III reviews the variabilities originating from the interface traps close to the graphene surface. These variabilities broadly exist in non-suspended graphene devices, and are of particular significance to evaluate the device stability. We discuss their physical principles, characterization methods, and the approach to minimize their effects via device engineering. For the second class, Section IV reviews the variabilities originating from the edge disorders within graphene itself, which represent the variation sources limited by material preparations. We examine the concepts of edge disorders, their effects on device performance, and ways of reducing them by improving the material quality. Section V reviews graphene variabilities from the metrology perspective. We present the possible use of the 'signal fluctuations' as novel probing mechanisms for sensing applications (e.g. probing the band structure, selective chemical sensing), and outlook their potential in nanometrology. Section VI concludes the paper with several further discussions.

## II. VARIABILITY SOURCES IN GRAPHENE

Graphene variabilities can be fundamentally viewed as sources/mechanisms that lead to deviations of the functional behavior from its ideal case [5]. We note that graphene can bear fluctuation mechanisms such as thermal noise, shot noise and electron-phonon/electron scatterings [83-86], which belong to the inherent properties of graphene and are out of the scope of our discussions.

Variability sources due to the non-ideality of graphene can be categorized into two broad classes (see Table I): I. Variabilities from the environmental disturbance are located close to the graphene surface and significantly affect the device properties (e.g. mobility, doping) [3, 64, 65, 87-90]. These variabilities are attributed to the external perturbations of the environment surrounding the graphene channel, such as the dielectric layer, the ambient environment, and the substrate. II. Variabilities from the imperfection of material quality are randomly distributed in graphene itself [91-95]. These variabilities are attributed to the geometrical variations of the

TABLE I. TYPICAL VARIABILITIES IN GRAPHENE. VARIABILITY SOURCES DUE TO THE NON-IDEALITY OF GRAPHENE CAN BE CATEGORIZED INTO TWO BROAD CLASSES: THOSE FROM THE ENVIRONMENTAL DISTURBANCE AND THOSE FROM THE IMPERFECTION OF MATERIAL QUALITY.

I) Variabilities from the environmental disturbance:
  1) Interface traps [65]
  2) Charged impurities [87, 88]
  3) Adsorbed molecules [89]
  4) Resonant scatterers [90]
  5) Surface roughness [3]
  6) Contact doping [64]

II) Variabilities from the material imperfection:
  1) Edge disorders [91]
  2) Structural defects away from the edge [92]
  3) Ripples [93]
  4) Curvatures [94]
  5) Other derivatives of graphene [95]

graphene material, posed by the limit of synthesis and preparation methods.

It is instructive to compare the above-mentioned variability sources in graphene with those in silicon technology [42, 43, 46, 96-99]. Both Class I and II graphene variabilities are analogous to the random variations in silicon devices (i.e. device parameter fluctuations in an unpredictable manner), however several differences exist between the sources of graphene and silicon variabilities. For example, some of Class I graphene variabilities (e.g. adsorbed molecules, surface roughness) are less problematic in silicon devices [43]. The variabilities near the graphene-dielectric/substrate interface have attracted considerable interest in graphene communities, mostly because they are 1) more influential in low-dimensional graphene devices with high surface-to-volume ratios [62, 71], and 2) relatively less understood than the Si-SiO$_2$ interface in CMOS technology [52]. Moreover, Class II graphene variabilities are the randomness specific to the handling of graphene materials, which are not exactly the same as those in silicon devices [43, 55, 100, 101]. For instance, edge disorders in graphene are not normally addressed in silicon devices possibly due to the maturity of CMOS technology. And the graphene-on-insulator or suspended graphene device structures are relatively immune from the random dopant fluctuation in the bulk silicon, one major random variation in CMOS processes. The differences between graphene and silicon variabilities need to be taken into account when integrating these two materials [40]. On the other hand, current research on graphene variabilities has focused on the device level up to the within-die and die-to-die scales [51, 57], whereas studies on the circuit and system levels in the wafer-to-wafer scale are still rare. Many systematic variations in silicon technology (i.e. definite spatial or temporal shifts caused by the tolerance of fabrication processes) would also be critical in the device engineering of graphene, such as length/width variations in lithography and etching steps, film thickness variations in deposition and growth processes, among others [42, 46, 99]. The related works would be crucial in evaluating the promise of graphene electronics.

Our discussions provide an overview of the similarities and differences between the variabilities in graphene and those in silicon. The nature of variability effects in graphene lies on its low-dimension and ways of preparing the material. The field is new, rapidly growing, and full of ample research opportunities. In the following, we will present the progress of two graphene variabilities, interface traps (Class I) and edge disorders (Class II), which have shown their critical roles in achieving high-performance graphene devices (Section III and IV) and opportunities for potential sensing applications (Section V).

### III. INTERFACE TRAP: GRAPHENE VARIABILITIES FROM ENVIRONMENTAL DISTURBANCE

Due to their large surface-to-volume ratio, graphene devices are sensitive to the interface traps close to the graphene surface [61, 71, 77]. They exist in the gate oxide or the substrate of non-suspended graphene devices, and can also be from the attached molecules or the surrounding environment [60-62, 70, 102]. Similar to silicon technology, trap-induced fluctuations in graphene pose a challenge to the device scalability since their effect increases as the device scales down [42, 103]. We next discuss their principles, characterization methods and how to minimize their effects via device engineering.

#### A. Device Fluctuations Caused by a Single Interface Trap

If a single trap close to the channel has its energy near the Fermi level (~$k_BT$, here $k_B$ is Boltzmann's constant and T is the absolute temperature), the device signal (e.g. current) will show two-level fluctuations in the time-domain (see Figure 1a) [103-106]. The random switching events (i.e. capture and emission of a single carrier from the channel by the trap) follow the Poisson statistics. In the frequency-domain, the power spectrum density (PSD) has a form of $S(f) = g \cdot \frac{\tau}{1+(2\pi f \tau)^2}$ with a corner-frequency and a pre-factor as $f_{corner} = 2\pi\tau = 2\pi(\tau_1^{-1} + \tau_2^{-1})^{-1}$ and $g = \frac{4\tau_1\tau_2}{(\tau_1+\tau_2)^2} \cdot (\Delta i)^2$ ($\Delta i$ is the current step in the time-domain; $\tau_{1,2}$ are the time constants of the two states), respectively [104, 106]. This Lorentzian-shape PSD is nearly constant when f<<$f_{corner}$, and approximately follows as $1/f^2$ when f>> $f_{corner}$ (see Figure 1b).

These two-level fluctuations belong to random telegraph noise (i.e. RTN, generation-recombination noise) [107-109]. It is commonly measured by sampling the current variations in the time-domain under a constant voltage bias across the channel (i.e. source-drain voltage) [105, 107]. RTN is a critical issue to the signal-to-noise ratio of devices due to its increasing effect with device scaling. However, RTN in graphene has not been systematically investigated at this stage, whereas many studies have been reported in carbon nanotube (CNT) [110-112]. The reason may be that the enclosed structure of CNT with a small diameter (2-5nm) can have fewer interface traps than a planar graphene sheet with a micron-sized width. RTN might be observable in small-area graphene devices at low temperature, such as a graphene nanoribbon with a nanometer-sized width [13, 14].

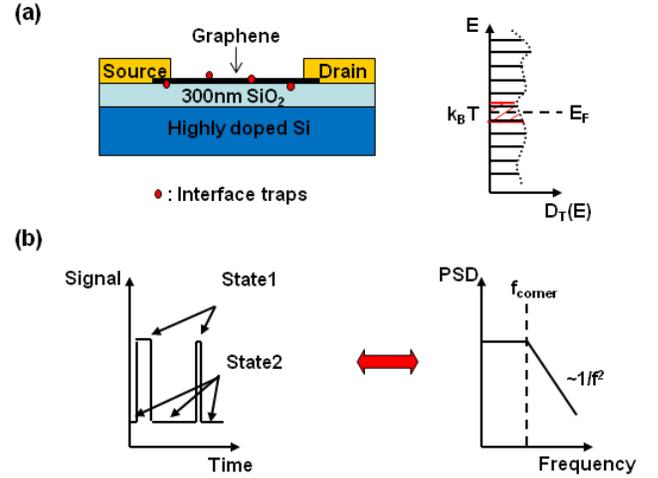

Fig. 1. **Device fluctuations caused by a single interface trap. a.** (Left panel) Schematics of a back-gated graphene device fabricated on a SiO$_2$/Si substrate. The interface traps (red dots) are located close to the graphene surface. (Right panel) Schematics of the density of trap states (D$_T$) in the energy scale (E). The traps with energy near the Fermi level (~ k$_B$T) can cause the random switching events, which lead to device fluctuations. **b.** (Left panel) Two-level signal fluctuations in the time-domain caused by a single interface trap (RTN). The time constants of the two states are labeled as τ$_{1/2}$. (Right panel) A typical Lorentzian-shape PSD of RTN in the frequency-domain. The PSD exhibits a corner frequency (f$_{corner}$) and a near 1/f$^2$ shape beyond f$_{corner}$.

#### B. Device Fluctuations from Many Interface Traps

Multiple-level fluctuations, on the other hand, are usually characterized by low-frequency noise measurement (LFN) in the frequency-domain (typically from 0.1Hz to 100 kHz) [113-115]. Although other physical mechanisms of LFN may co-exist [109, 116, 117], interface traps are usually employed to understand the origin of LFN in graphene devices [61, 65, 102, 118, 119]. For instance, suspended graphene devices show a 6-12 times lower LFN than those with a SiO$_2$ substrate, suggesting that the traps in SiO$_2$ substrate contribute significantly to LFN [118]. On the other hand, recent studies have suggested that the LFN in graphene depends on various scattering mechanisms, the environment near the graphene surface and the sample qualities [65, 77, 102, 119]. For instance, Heller et al. [102] have proposed that the LFN behavior in liquid-gated graphene devices can relate to the charge-induced local potential fluctuations near the graphene-electrolyte interface (i.e. the charge noise).

The McWhorter model views the LFN as the multi-level fluctuations caused by an ensemble of many interface traps (number >>1): each trap contributes a RTN over a wide range of $\tau$ and $f_{corner}$ (see Figure 2a) [109, 117, 120, 121]. The overall PSD can be integrated as $S(f) = \int_\tau g \cdot \frac{\tau}{1+(2\pi f \tau)^2} \cdot p(\tau) d\tau$, where $p(\tau)$ is the distribution function of $\tau$. Assuming that i) electrons reach the traps by tunneling and τ depends exponentially on the distance from the channel (z), one has $\tau = \tau_0 \exp(z/\lambda)$ (λ is the penetration depth), and ii) traps are uniformly distributed along the z-direction (i.e. $\frac{dN_T}{dz} = const$ with N$_T$ as the number of traps) [113, 117], one has $p(\tau) \sim \dfrac{dN_T}{d\tau} = \dfrac{dN_T}{dz} \cdot \dfrac{dz}{d\tau} = \dfrac{dN_T}{dz} \cdot \dfrac{\lambda}{\tau}$ and $S(f) \propto \int_\tau \dfrac{1}{1+(2\pi f \tau)^2} d\tau \propto \dfrac{1}{f}$, which is known as the 1/f noise. The McWhorter model is physically intuitive and popular in LFN theorem, but it only holds when LFN is dominated by the carrier number fluctuations [113, 117, 121]. At present, it is still unclear about the relative contribution of carrier number fluctuations to the overall LFN in graphene devices.

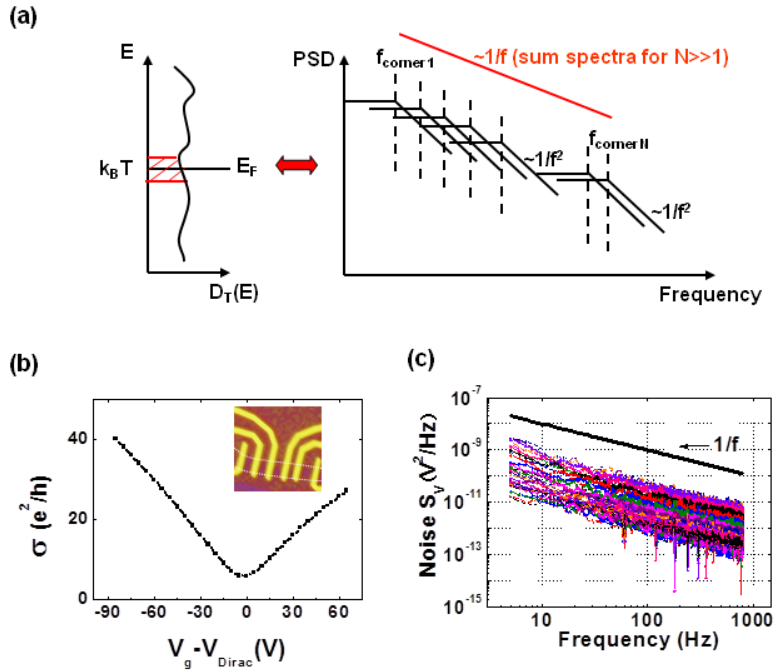

Fig. 2. **Device fluctuations from many interface traps. a.** Schematics of the McWhorter model (Left panel). If there are many interface traps with energy close to Fermi level (several $k_B T$), all of them will contribute to the carrier trapping/detrapping processes. (Right panel) An ensemble of many trap-induced RTNs with a wide range of time constant and corner frequency ($f_{corner}$) can result in LFN with a 1/f shape of PSD. **b.** The dependence of DC conductivity ($\sigma$) on the gate bias ($V_g$) measured in a single-layer graphene device under four-probe configuration (shifted by the gate bias at the Dirac point, $V_{Dirac}$) [65]. The inset shows the optical image of a fabricated multi-terminal device (graphene was outlined in dotted lines). **c.** Typical room-temperature LFN spectra of a back-gated graphene sample [65]. The four-probe noise spectra ($S_V$) followed $1/f^\alpha$ behavior with $\alpha$ ranging from 0.85 to 1.12 with gate biases varying from -50V to 100V.

The importance of LFN for practical applications stems from the fact that it contributes to the phase noise of the devices and systems via unavoidable non-linearity. A low-level phase noise is a critical requirement for high-frequency applications of graphene [61, 122]. LFN measurements are broadly used as a characterization technique to provide information about interface traps in graphene devices [60, 61, 63, 65, 77, 123]. Figure 2b and 2c show a typical room-temperature LFN measurement of a back-gated single-layer graphene (SLG) device [65]. The study employed a four-probe configuration to minimize the noise contribution from the contacts (see the inset of Figure 2): an Agilent 4156C was used to apply dc current bias to the device, and measure its dc conductivity $\sigma$; an Agilent 35670A was used to collect the noise spectra of the fluctuations of the potential difference (V) across the graphene samples. At each gate bias (shifted by $V_{Dirac}$, the gate bias at Dirac point), the conductivity was averaged 10 times at the same time of the noise measurement in order to ensure the data consistency (see Figure 2b). The noise spectra ($S_V$) were averaged 20 times from the fast Fourier transforms of the sampled voltage signal (V), and subtracted by the background measured at zero current bias (see Figure 2c). The measured LFN followed the 1/f shape (i.e. $1/f^\alpha$, $\alpha \sim 1$) at each gate bias, whereas the deviation may relate to pronounced Lorentzian spectra from individual RTNs or other noise sources.

To quantify the data, we can normalize LFN as $(\dfrac{S_I}{I^2})_{V=const} = \dfrac{S_G}{G^2} = \dfrac{S_R}{R^2} = (\dfrac{S_V}{V^2})_{I=const} = \dfrac{A}{f^\alpha}$ by assuming the resistance/conductance (R/G) is independent from the bias condition [109]. The parameter A is commonly used as a measure of the LFN amplitude [61, 65, 115, 124]. Alternatively, the Hooge parameter $\alpha_H$ is empirically defined as $(\dfrac{S_I}{I^2})_{V=const} = \dfrac{A}{f^\alpha} = \dfrac{\alpha_H}{f^\alpha N}$ (N is the total number of carriers), whose value depends on the details of the materials and devices [116, 125, 126]. As such, Table II summarizes the typical LFN measurements in graphene and graphene nanoribbon (GNR) devices [61, 63, 65, 70, 77, 123]. The measured $\alpha_H$ value (from $10^{-5}$ to $10^{-2}$) depends on many factors such as contact material, device environment and sample quality. For example, the noise in chemical-vapor-deposition (CVD) based graphene devices ($\alpha_H \sim 8-9 \times 10^{-3}$) can be comparable to that in exfoliation-based graphene devices ($\alpha_H \sim 10^{-3} - 10^{-2}$) [65, 70, 77]. This result suggests a low amount of impurities left on graphene during the CVD growth and transfer processes (much better than previous results [127]). On the other hand, bilayer GNRs (BLR) were reported to exhibit a lower noise than single-layer GNRs,

TABLE II. TYPICAL $\alpha_H$ VALUES IN GRAPHENE AND GNR DEVICES AT ROOM TEMPERATURE. THE MEASURED $\alpha_H$ VALUE (FROM $10^{-5}$ TO $10^{-1}$) DEPENDS ON FACTORS SUCH AS CONTACT MATERIAL, DEVICE ENVIRONMENT AND SAMPLE QUALITY. THE LFN DATA WERE MEASURED IN EITHER DUAL-GATED [63] OR BACK-GATED DEVICES [61, 65,70,77,123]. ABBREVIATIONS: SINGLE-LAYER GRAPHENE (SLG); BI-LAYER GRAPHENE (BLG); FEW-LAYER GRAPHENE (FLG); MULTI-LAYER GRAPHENE (MLG); SINGLE-LAYER GNR (SLR); FOUR-PROBE MEASUREMENT (4T).

| Data source | Contact | $L_{channel}$ (μm) | Exposure | Sample | Typical Mobilities (cm$^2$/(Vs)) | $\alpha_H$ (away from Dirac point) |
|---|---|---|---|---|---|---|
| G. Liu [63] | Cr/Au (2T) | 9 | Ambiance + HfO$_2$ TG | SLG | $\mu_e$~1550 $\mu_h$~2200 | $2 \times 10^{-3}$ |
| Q. Shao [123] | Cr/Au (2T) | 9 | Ambiance | BLG | $\mu_{e/h}$~3000 | ~$1 \times 10^{-4}$ |
| A. N. Pal [70] | Au (4T) | ~2-4 (Fig. 1b) | Vacuum | SLG/FLG/MLG | ~1100 (SLG), ~2450(FLG), ~1200(MLG) | $2 \times 10^{-2}$ (SLG), $5\text{-}8 \times 10^{-5}$ (FLG /MLG) |
| Y. M. Lin [61] | Pd (2T) | 0.8-2.8 | Vacuum | SLR/BLR | ~700 (SLR) | ~$10^{-3}$ (SLR) |
| G. Xu [65] | Ti/Au (4T) | 1.5-2.8(SLG) 1.9-5.5(BLG) | Vacuum | SLG/BLG | ~2000-4000(SLG) ~1000-5000(BLG) | ~$10^{-3}$-$10^{-2}$ (SLG/BLG) |
| A. N. Pal [77] | Au (4T and 2T) | 15 | Vacuum | CVD-based SLG | ~400 at room temperature | $8$~$9 \times 10^{-3}$ |

which can relate to their different band structures and transport properties [61, 128].

*C. Approach to Minimize the Effect of Interface Traps*

Trap-induced fluctuations in graphene degrade the device performance, forming a critical issue that needs to be addressed. Many efforts are made to minimize their effect via device engineering, which include device passivation[63], the use of multilayer graphene channels [64, 70], surface cleaning [60] and substrate engineering [3, 118].

A recent study has reported the measurements of LFN in the double-gated graphene devices, which helped to understand the effect of passivation on the noise level [63] (see Figure 3a). Single-layer graphene sheets (width~10μm) were exfoliated from highly oriented pyrolytic graphite onto a thermally-grown 300nm SiO$_2$ dielectric layer on a highly-doped Si substrate which acts as the backside gate (BG). A 20nm HfO$_2$ layer achieved by atom-layer deposition (ALD) method was used as the top-gate dielectric, which was patterned by electron-beam lithography to partially cover the graphene channel. The Cr/Au metal layers (5/60nm) were evaporated to serve as the source, drain and top-gate electrodes. A two-terminal LFN measurement was conducted in ambience by monitoring the PSD of the current ($S_I$) with a constant drain-to-source bias ($V_{ds}$). The devices benefit from the top-gate passivation by the HfO$_2$ layer, where the gated-channel is immune to the possible traps from ambiance. The measured LFN was low ($\alpha_H$ ~$10^{-3}$), and found to be mainly due to the un-gated graphene channel (i.e. uncovered by the HfO$_2$ layer). This result suggests that a low noise graphene device can be achieved by improving the passivation of the channel.

Using multilayer graphene (i.e. >1 layer) as the channel material is another approach to achieve low noise devices. For example, a graphene sheet with more than three layers (FLG) has been found to have much lower LFN than that in single-layer graphene (SLG), which are of practical interest [70]. The physics can be explained as the efficient screening to the interface traps in FLG with more layers, whereas the difference in the band structures between FLG and SLG also plays a role [66, 70]. In this context, the thickness-graded graphene transistor (GTG) was utilized to study the layer dependence of LFN [64]. The fabrication method was the same as in the conventional back-gated graphene devices. The channel of GTG devices was confirmed to gradually vary from a single-layer in the middle to multilayers ($\geq$ 2 layers) near the source/drain contacts. The measured LFN in GTG and BLG was typically lower than that in SLG with the same device area (S) (see Figure 3b). The lower LFN in GTG and BLG (than that in SLG) can relate to a smaller metal-doping induced Fermi-level shifts near the contact and/or more effective carrier screening to the interface traps, which result from the difference in their band structures and transport properties [61, 64]. This result shows the layer dependence on the LFN in graphene, providing insight for designing low-noise devices with multilayer graphene. The contact engineering is yet another significant aspect in lowering the LFN in graphene devices. However no consensus about the weight of LFN contribution from the metal-graphene contact (compared to that from the graphene channel) has been reached yet [64, 65, 77, 119]. The results can relate to the fraction of the contact resistance over the entire device resistance in as-made graphene devices, which depends on the device qualities and the contact material in use [115, 123].

Cleaning the graphene surface is also helpful in lowering the effect of interface traps. The adsorbed particles left on graphene surface during the fabrication, for example, can significantly affect the device performance [129-131]. Xu *et al.* [60] have reported an improved LFN level in GNR devices by cleaning the graphene surface. As shown in Figure 3c, GNRs were patterned using a Si-nanowire mask to avoid the use of photoresist (e.g. HSQ) which strongly affects device performance [129]. A 20min, 100°C vacuum annealing process was conducted to desorb the contaminants on the graphene surface. The cleaned single-layer GNR devices feature an improved hysteresis (during a dual sweep of the gate bias) and a 30% lower LFN than those achieved by HSQ-based methods [61]. The LFN improvement can be attributed to a cleaner graphene surface, whereas the four-probe configuration also reduces the noise contributed from the contacts as in two-probe setups. This work presents an approach of achieving low-noise GNR devices using the nanowire-patterning method.

Lastly, improving the substrate quality can also reduce the number of interface traps in graphene devices. For instance, boron-nitride (BN) substrate shows its potential in achieving high-performance graphene devices, which outweigh those on SiO$_2$ substrates [3, 40]. Its superior properties, such as ultra-flat surface with very few dangling bonds and charge traps, are expected to minimize the trap-induced fluctuations and lower the noise. Suspending graphene (i.e. free of substrate) is another option to greatly reduce the LFN with much fewer traps in devices [119], which may show promise for low-noise applications. A recent study shows that the graphene devices can have 6-12 times lower LFN by etching away the $SiO_2$ substrate, which can benefit the device performance in pH-sensing applications [118].

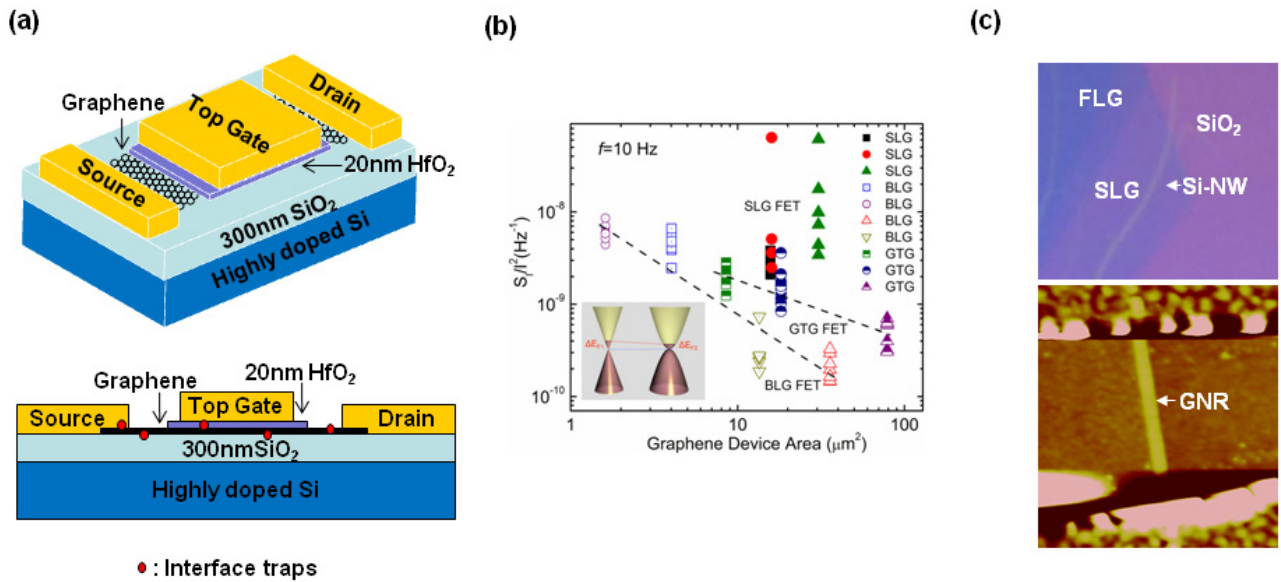

Fig. 3. **Device engineering to minimize the effect of interface traps in graphene devices. a.** (Top panel) A dual-gated single-layer graphene (SLG) device with a $SiO_2$/Si substrate (back-gate) and a 20nm $HfO_2$ layer as the top-gate dielectric [63]. (Bottom panel) The low-noise SLG devices ($\alpha_H \sim 10^{-3}$) benefit from the top-gate passivation, where the noise was mainly from the un-gated graphene channel. **b.** Comparison of the normalized LFN values ($S_I/I^2$ at f=10Hz with |V$_g$-V$_{Dirac}$|≤30V) in thickness-graded graphene (GTG), single-layer graphene (SLG) and bilayer graphene (BLG) devices [64]. The lower noise in GTG (than that in SLG) can be from the reduced noise contribution near the contact with more layers. **c.** (Top panel) The optical image of the silicon nanowires patterned onto single layer (light blue) and multilayer (dark blue) bulk graphene sheet on top of 300nm $SiO_2$ layer (purple) [60]. (Bottom panel) The atomic-force-microscopy (AFM) image of a typical BLR after removal of the nanowire-mask (Ti/Au contacts). This fabrication method avoided the photoresist contamination (e.g. HSQ) to the GNR surface, which helps lower the noise.

Reducing the effect of interface traps is a key issue to continued miniaturization of graphene devices, circuits and systems. Future studies require the clear understanding of their physical mechanisms and a systematic optimization of the device engineering.

## IV. EDGE DISORDER: GRAPHENE VARIABILITIES FROM MATERIAL IMPERFECTION

Unlike a CNT with a perfectly enclosed structure, graphene usually has unavoidable edge disorders for its planar geometry [12]. As the width of graphene narrows down to the nanometer scale, a graphene nanoribbon (GNR) would become very sensitive to the scattering induced by these edge disorders [5]. One should also mention that the edge disorder produces a major detrimental effect on the graphene thermal conductivity [132-135]. The question of how to well control the edge disorders in GNR devices is essential in evaluating the graphene scaling along the width direction and the practicality of GNR electronics – GNR has an energy gap that benefits the device switching; however, its mobility can be seriously degraded by edge disorders[1, 5]. We next discuss their concepts, roles on device performance, and the ways of reducing their impact by improving the material quality.

### A. Type of Edge Disorders

Various edge disorders exist in as-made GNR devices [91]. It is important to identify the difference in their origin, morphology and length-scale. Here we discuss the main categories of edge disorders that are commonly referred to in the literature (see Figure 4):

i) Carbon atoms (C-atom) form dangling bonds at the GNR edge, which can bind to different atoms, including H, O, F and OH (see Figure 4a)[136]. Due to the difference between the local density of states at the edge and that in the center of the GNR, these edge disorders serve as scattering sites. This type of edge disorders can even exist in GNRs with perfect zigzag or armchair edges.

ii) Mixed edge structure composed of both zigzag and armchair edges are broadly observed in as-made GNRs (see Figure 4b)[12, 137], whereas edge structures beyond zigzag and armchair edges are also reported [138]. The existence of these edge disorders partly explains why the chirality dependence of GNR can be diluted in real samples [139], which do not follow the theoretical predictions based on pure zigzag or armchair edges.

iii) C-atoms at the edge can restructure themselves into other morphologies (see Figure 4c). These edge disorders can be in the form of dislocations within or out of the GNR plane [91, 140]. Examples include point defect, vacancies, 5-7-5 or 5-8-5 edge deformations, loops, line-defects, adatoms and

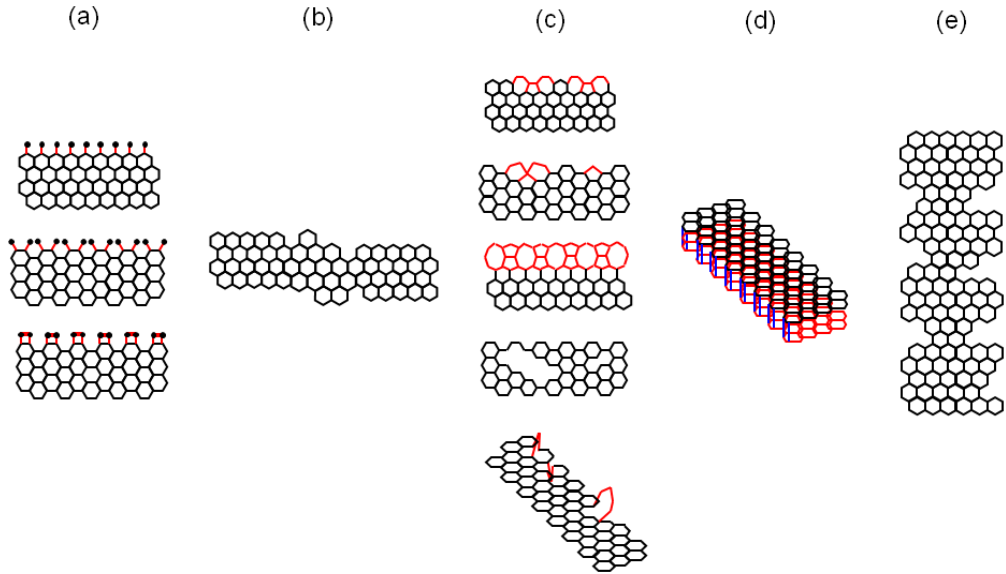

Fig. 4. **Main categories of edge disorders in as-made GNR/graphene materials. a.** The dangling bonds of the outmost C-atoms can bind to H/F/OH/O atoms (in red) [136], which even exist near the perfect zigzag (top panel) or armchair edges (middle and bottom panels). The top and middle panels show typical H or F atoms near the GNR edge, where the bottom panel shows typical O atoms near the GNR edge. **b.** Mixed edge structure composed of both zigzag and armchair edges [12, 137]. **c.** Restructured C-atoms near the edge (in red) [91, 140]. Examples are illustrated from top to bottom: (In-plane) 5-7-5 dislocations near a zigzag edge, 5-7-5 dislocations near an armchair edge, 5-8-5 dislocations near a zigzag edge, single/multiple vacancies, (out-of-plane) adatoms/interstitials. **d.** Enclosed edges (in blue) in bilayer or multi-layer graphene families (number of layers>2) [141-143]. Note: the blue part is just for illustration - C-atoms from different layers may not connect this way. **e.** Line edge/width roughness (LER/LWR) [59]. They have a spatial size of $10^0 \sim 10^1$ nm in the nanowire-mask based GNR devices.

interstitials. Study of this type of edge disorders is at its early stage.

iv) Multilayer GNRs can form a partially-closed edge structure (see Figure 4d). This type of edge disorders has been found in both GNRs and micron-wide graphene sheets [141-143]. They may weaken the edge-induced carrier localization in multilayer GNRs compared to that in single-layer GNRs [144], since the carriers can couple among different layers through the closed edges.

v) Line edge roughness (LER) and line-width roughness (LWR) also exist in as-made GNRs (see Figure 4e). Depending on the material preparations, LER/LWR can have a length-scale of $10^0$-$10^1$nm, which is directly observable using atomic-force-microscopy [59]. If LER/LWR becomes serious, GNR behaves more like a chain of connected quantum dots and forms a Coulomb blockade [145]. LER/LWR is expected to be more controllable by material engineering than the edge disorders with an atomic scale.

Overall, the edge disorders in 1) to 3) can be treated as local disturbances (short-range scatters) near the GNR edge; those in 4) and 5) relate to the irregularities that deviate from ideal GNRs [91, 140]. Whichever categorizations we take, it is clear that the component of edge disorders in GNR (and graphene sheet) is complicated. Due to the lack of accurate manipulation of these edge disorders, most experiments could not differentiate the role of a certain type of edge disorders from the others at the moment. The solution can appear with the advancement of fabrication and characterization methods.

*B. Effects of Edge Disorders on Device Performance*

Edge disorders in as-made GNRs raise the concern of their impact on transport properties and device operations. For example, the transport gap observed in GNRs makes them advantageous in switching on/off the devices, while the question of how the edge disorders affect or contribute to the observed gap in GNR devices is still debated [130, 146, 147]. Han *et al.* [130] have reported the size-dependence of single-layer GNR (SLR) at low carrier densities, attributing the transport gap to a combination of the edge effect and the Coulomb charging effect. However, their fabrication method leaves chemical residues (HSQ) on top of the GNR samples, which makes it difficult to probe the intrinsic GNR properties [139]. In contrast, a study of SLRs prepared by a metal-mask etching method suggests that the transport gap mainly originates from the effect of charged impurities instead of edge disorders [147]. Given the sensitivity of GNRs to the weight of various scatterings, it may not be surprising to see the inconsistency of the role of edge effects in these measurements, since SLR devices fabricated through different methods can yield quite different transport properties (see details in Table III) [13, 14, 60, 129, 139, 147-157]. Details in sample preparations can affect the weight of edge-induced scattering in specific GNRs, and need to be considered for comparing the results. On the other hand, Yang and Murali [150] have studied the width-dependence of carrier mobilities in GNR-array devices patterned by lithography. The decrease of mobility in samples with a narrower width was attributed to the increased scattering from LER, which poses a constraint to the device performance.

Xu *et al.* [58] have recently employed the length-dependence of sample resistance (i.e. resistance scaling, R-L relation) to investigate the role of edge disorders in the transport of single-layer and bilayer graphene (SLG, BLG) and GNR (SLR, BLR) devices (see Figure 5). According to one-parameter scaling law, R-L relation can identify the transport regimes in low-dimensional systems, such as the exponential R-L relation and linear R-L relation for localization and diffusion regimes, respectively [81, 158, 159]. Here GNRs were fabricated by a nanowire-mask etching method with good performance as reported before [60]. The room-temperature sample resistance was measured within the low-bias regime at both low and high carrier densities. The experimental data showed that the SLR transport lies in a strong localization regime (exponential R-L relation), which can be attributed to a strong edge effect (see Figure 5a). In contrast, BLRs featured diffusive transport (linear R-L relation), where the absence of localization can relate to a weaker edge effect than that in SLRs (see Figure 5b).

TABLE III. Typical fabrication methods of GNR devices. Plasma-based GNR fabrication methods typically use $O_2/Ar$ plasma etching to form the GNR, with chemical resists, nanowires or metal lines being employed as the etching mask [14, 60, 129, 139, 147-151, 154-157]. On the other hand, GNRs can be chemically-derived via unzipping CNTs [152, 153] or thermally-grown on SiC wafers [13]. The chemical-solution based method can achieve ultra-narrow GNRs with a sub-5nm width [152]. Since GNR is sensitive to multiple types of scattering, it may not be surprising to see the inconsistency in their transport properties, which can be quite different in samples prepared by different methods. HSQ and PMMA are the photoresist materials used in ebeam-lithography.

| Data source | Contact | GNR preparation | GNR surface |
|---|---|---|---|
| M. Y. Han [139] | Cr/Au | HSQ mask + $O_2$ plasma | HSQ coverage |
| Y. M. Lin [129] | Pd | HSQ mask + $O_2$ plasma | HSQ removed by HF |
| C. Lian [148] | Cr/Au | Al mask + $O_2$ plasma | Al removed by etchant |
| P. Gallagher [147] | Ti/Au | Ti mask + $O_2$ plasma | Ti and PMMA removed |
| J. Bai [149] | Ti/Au | Si nanowire mask + $O_2$ plasma | Si nanowire coverage |
| Y. Yang [150] | Ti/Au | HSQ mask + $O_2$ plasma | HSQ coverage |
| Z. Chen [151] | Pd | HSQ mask + $O_2$ plasma | HSQ removed by HF |
| L. Jiao [152] | Pd | gas phase oxidation + sonication | GNR formed by unzipped CNT |
| L. Jiao [153] | Pd | PMMA protection + Ar plasma | GNR formed by unzipped CNT + PMMA removed |
| G. Xu [60] | Ti/Au | Si nanowire mask + $O_2$ plasma | Si nanowire removed |
| M. Sprinkle [13] | Pd/Au | SiC growth on $(1\bar{1}0n)$ facet (n~8) | GNR formed by SiC |
| R. Yang [154] | Ti/Au | PMMA mask + $O_2$ plasma + $H_2$ plasma etching | PMMA removed |
| X. Wang [155] | Pd | Al mask + Ar plasma + ($NH_3$ + $O_2$) etching | Al removed by etchant |
| X. Li and X. Wang [14] | Pd | Chemically-derived from solution | GNR dispersed on substrate |
| D. V. Kosynkin [156] | Pt | Chemically-derived from solution | GNR formed by unzipped CNT |
| J.B. Oostinga [157] | Ti/Au | PMMA mask + Ar plasma | PMMA removed |

Through the comparisons among SLR, BLR, SLG and BLG, the edge effect in graphene materials was found to be reduced by enlarging the width, decreasing the carrier densities or adding an extra layer (see Figure 5c). From SLR to SLG, the data showed a dimensional crossover of the transport regimes possibly due to the drastic change of the edge effect. These results reveal a critical role of edge effect in graphene transport and thus the resistance scaling rules, which may provide insight to achieve scalable graphene electronics.

Another recent work has reported a direct analysis of the LWR in GNR devices fabricated by the nanowire-mask method [59, 160]. The edge profile of GNRs was extracted from their AFM/SEM images by an image-processing algorithm (see Figure 6a). Then, the width values were sampled along the L-direction of the edge-profile image, which were used to calculate the standard deviation ($\sigma$) as the LWR amplitude. The LWR amplitude among 13 SLRs and 5 BLRs was found to generally decrease with the GNR width (W), and the smallest LWR amplitude was below 5nm for SLRs with W~30nm [59]. This result can relate to the etching undercut due to the circular cross-section of the nanowire-mask. The W-dependence of on/off ratios in the GNRs with different $\sigma$ values was measured to evaluate the LWR impact on device performance (see Figure 6b, the $G_{on}/G_{off}$ ratio at T=300K is calculated by the measured conductance (G) at $|V_g-V_{Dirac}|=30V$ and $V_g=V_{Dirac}$, respectively). The data showed a large variation in the W-dependence of the $G_{on}/G_{off}$ (e.g. the $G_{on}/G_{off}$ value in SLRs varies from 2.2 to 3.5 near W~40nm), with no clear dependence on the $\sigma$ values. This large variation of on/off ratios in GNRs in the presence of LWR is consistent with theoretical works [161-163]; however it may not be fully attributed to LWR because other atomic-scale edge disorders can also contribute to the variations. Although LWR itself could lead to device degradation, the complexity in the component of edge disorders needs to be taken account in as-made GNRs.

### C. Advancement in Reducing Edge Disorders in Graphene

Edge disorders represent the graphene variabilities posed by material preparations, which challenge the reliability and scalability of graphene systems. Much progress has been made to reduce their effect through the advances of material synthesis and patterning methods, with some technology showing very promising results (see details in Table III). For example, Wang *et al.* have developed a gas phase etching chemistry to narrow the GNRs down to <10 nm with a well-controlled etching rate [155]. The achieved sub-5nm-wide GNR devices show a high on/off ratio up to ~$10^4$ at room temperature. On the other hand, Jiao *et al.* [153] have demonstrated high-performance GNRs derived from unzipping multi-walled CNT samples. They achieve this by either plasma etching of the CNTs in a polymer film, or mechanical sonication of the gas-phase oxidized CNTs in an organic solvent. The obtained GNR devices have very smooth edges and the room-temperature mobility as high as 1500cm$^2$/(Vs) with a 10-20nm width. A recent study shows that these GNRs can behave as perfect quantum wires under low temperature [15]. Moreover, Cai *et al.* [164] reported an atomically precise bottom-up fabrication of GNRs, which can provide GNRs with engineered chemical and electrical properties. The topology, width and edge profile of GNRs are well-defined by the structure of the precursor monomers, which are prepared by surface-assisted coupling and cyclohydrogenation. This work provides another approach in developing high-performance GNR devices. Overall, we expect that the continuous technology improvement with an accurate manipulation of the edge disorders will bring much more excitement in graphene community.

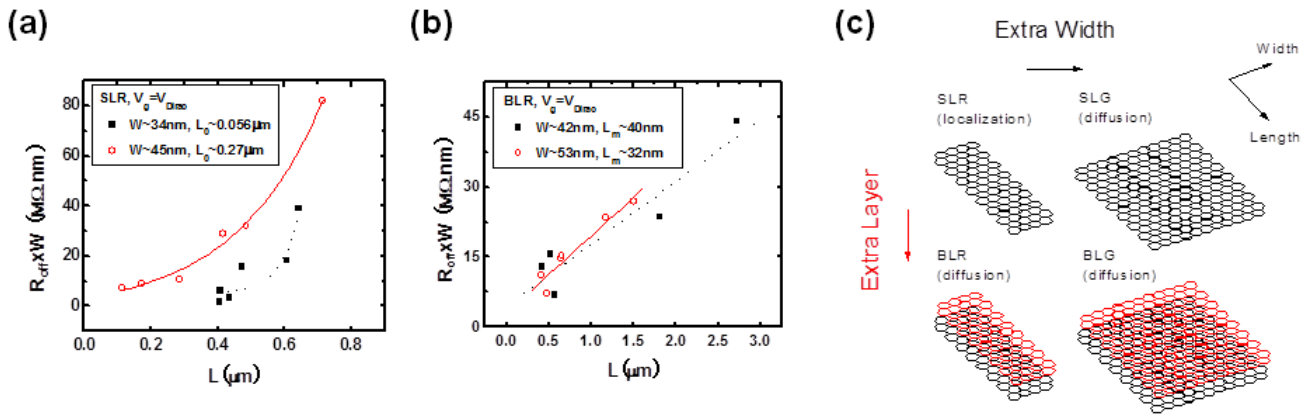

Fig. 5. **Effect of edge disorders on resistance scaling rules in graphene nanostructures [58]. a.** Room temperature R-L relations for SLR at off-state ($V_g = V_{Dirac}$), where $R_{off}$ exponentially increases with L. The fitting showed a characteristic localization length of $L_0 \sim 0.27\mu m$ for W~45nm and $L_0 \sim 0.056\mu m$ for W~34nm, respectively. **b.** Room temperature R-L relations for BLR at off-state ($V_g = V_{Dirac}$), where $R_{off}$ linearly increases with L. The fitting showed a characteristic mean-free-path of $L_m \sim 40nm$ for W~42nm and $L_m \sim 32nm$ for W~53nm. **c.** Schematics for the crossover of transport regimes in graphene devices. This edge effect in SLR can be weakened by either adding an extra layer to form BLR or increasing the width to form SLG; both cause the transition of transport regimes from localization to diffusion.

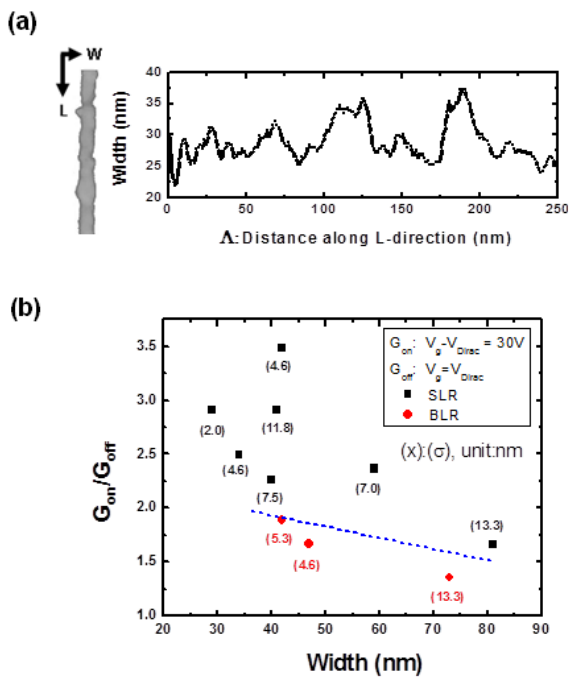

Fig. 6. **Line-width roughness (LWR) analysis for GNRs [59]. a.** (Left panel) The extracted GNR edge profile from the atomic-force-microscopy image for LWR analysis. (Right panel) The width sampling (smoothened) along the L-direction of a single-layer GNR versus the distance along the L-direction ($\Lambda$). The LWR amplitude, $\sigma$, was defined as the standard deviation of the sampled width values. **b.** The on/off ratios ($G_{on}/G_{off}$) of as-made GNRs (both SLRs and BLRs, T=300K) versus the averaged width (W). The low-bias conductance ($G_{on}/G_{off}$) at both on- and off-states were measured at $V_g - V_{Dirac} = 30V$ and $V_g = V_{Dirac}$, respectively. The values of $\sigma$ were labeled at each data point (unit: nm). The guide to the eye showed that the on/off ratios are generally lower in BLRs than those in SLRs.

## V. GRAPHENE VARIABILITIES FOR SENSING APPLICATIONS

While variabilities in graphene are generally considered the challenges for device engineering, a re-visit of their concepts can open up potential applications. Recent studies have revealed strong correlations between the 'signal fluctuations' in graphene devices with their surrounding environment and the material properties (e.g. band structures) [65, 71, 75, 76]. By characterizing the graphene variabilities (e.g. LFN), one can thus probe the environmental change near the graphene surface and the alteration of graphene properties. This variability-based probing mechanism, although at an early stage of development, can be useful in graphene-based sensing applications. Similar ideas have been implemented in silicon-nanowire and carbon nanotube devices, where the LFN can be employed in gas sensing and biosensing with high sensitivities [82, 165].

For example, charged impurities left near the graphene-SiO$_2$ interface can create an inhomogeneous charge distribution along the graphene sheet, which is a dominating scattering mechanism that limits the carrier mobility and can be responsible for several physical anomalies near the Dirac point (see Figure 7a) [5, 87, 88, 166, 167]. To investigate how the presence of spatial charge inhomogeneity influences the LFN behavior in graphene, Xu *et al.* [65] conducted research on the gate-dependence of the LFN amplitude (A) in back-gated SLG and BLG devices built on a SiO$_2$/Si substrate. Graphene devices were maintained in a vacuum environment and a 20min vacuum bakeout (100°C) process was generally applied before the LFN measurements. Using a four-probe measurement setup as described before (see Section II), the gate-dependence of LFN in SLG and BLG was found to feature an M-shape and V-shape, respectively (see Figure 7b and 7c). The analysis showed that the noise behavior near the Dirac point can be attributed to the extent of spatial charge inhomogeneity at low carrier density limits (e.g. the noise maximum in the M-shape of SLG matches the density of charged impurities, $n_{imp}$). The correlation between the gate-dependence of LFN and the spatial charge inhomogeneity in graphene can act as a probing mechanism to characterize the non-uniform doping profile of graphene. For instance, the LFN spectroscopy indicates the amount of charged impurities near the graphene surface, which can be used to evaluate the substrate/dielectric quality of graphene devices.

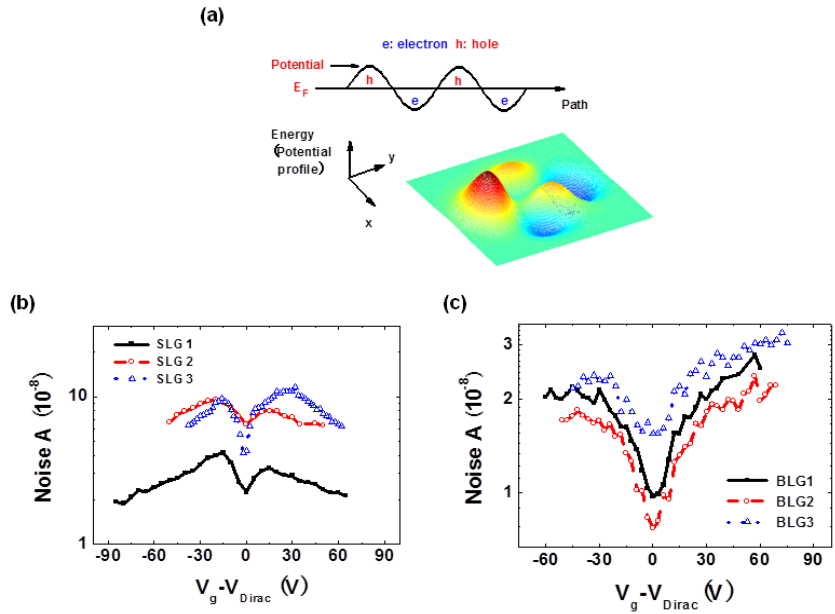

Fig. 7. **Gate-dependence of LFN in graphene devices: Probing the charge impurities by noise behavior [65]. a.** Charged impurities near the graphene-$SiO_2$ interface can create an inhomogeneous charge distribution along the graphene sheet, which is a dominating scattering mechanism that limits the carrier mobility. **b.** Gate-dependence of LFN amplitude (A) in SLG featured an M-shape at room temperature (shifted by the gate bias at the Dirac point, $V_{Dirac}$). **c.** Gate-dependence of LFN amplitude (A) in BLG featured a V-shape at room temperature (shifted by $V_{Dirac}$). The analysis showed that the noise behavior near the Dirac point can be correlated to the extent of spatial charge inhomogeneity at low carrier density limits (e.g. the noise maximum in the M-shape of SLG matches the density of charged impurity, $n_{imp}$).

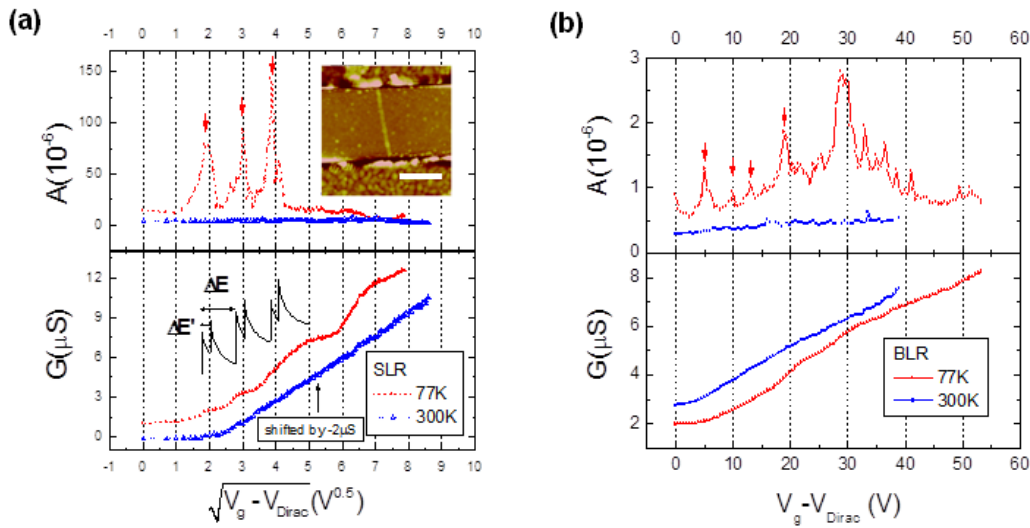

Fig. 8 **Gate-dependence of LFN in GNR devices: Probing the band structure by noise behavior [71]. a.** Temperature-dependent noise (A, top) and conductance (G, bottom) in SLR (W~42 nm, L~0.81 μm) in the scale of $\sqrt{V_g - V_{Dirac}}$ for the electron-conduction side ($V_g > V_{Dirac}$). The upper inset shows the AFM image of the SLR device with a scale bar equal to 0.5μm. At 77K, the gate-dependence of LFN showed peaks whose positions quantitatively matched the subband positions in the quasi-1D DOS (see the lower inset for the schematics). The strong correlation between LFN and DOS provides a robust mechanism to electrically probe the band structure of GNRs; **b.** Temperature-dependence of LFN behavior in BLR (W~49 nm, L~0.79 μm) was presented in scale of $V_g - V_{Dirac}$ for the electron-conduction side ($V_g > V_{Dirac}$). At T=77 K, the noise peaks (as arrowed) appeared while the corresponding conductance plateaus (bottom) were not obvious.

Taking one step further, Xu and co-workers [71] extended the LFN study in back-gated SLR and BLR devices, aiming to investigate the impact of their quasi-one-dimensional (1D) transport on the noise behavior. The GNR devices were achieved by the nanowire-mask based method (see the inset of Figure 8a), and kept in vacuum for both LFN and DC conductance measurements. Data were presented in the energy scale to compare with the band structure ($\sqrt{|V_g - V_{Dirac}|}$ in SLR, $|V_g - V_{Dirac}|$ in BLR). Through the analysis, the enhanced conductance fluctuations (noise) were found to originate from the quantum confinement along the GNR widths. In SLRs, the gate-dependence of LFN showed peaks whose positions quantitatively match the subband positions in the quasi-1D band structures (see Figure 8a, W~42nm). The LFN peaks can be attributed to the enhanced trap-induced conductance fluctuation: the fluctuating occupancy of interface traps causes the variation of the local potential; the resulting conductance fluctuation is enhanced near the subband thresholds where the density-of-states (DOS) diverges [71, 79]. In BLRs, the LFN peaks were also obvious while the subband feature was unclear in conductance data (see Figure 8b, W~49nm). Overall, the correlation between LFN and DOS provides a more robust mechanism to electrically probe the band structure of GNRs than the conductance measurement (where the subband feature is not clearly seen from the conductance plateaus).

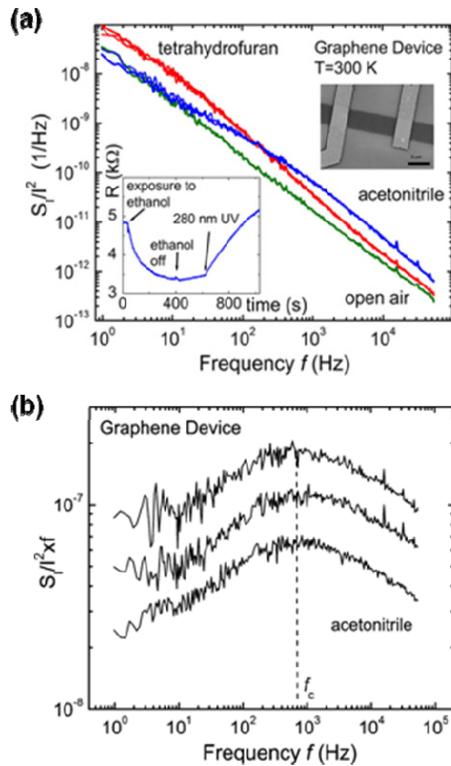

Fig. 9. **Selective gas sensing in back-gated graphene devices [75] a.** LFN spectra of SLG devices measured in open air and under the exposure to acetonitrile and tetrahydrofuran vapors (T=300K at $V_g$=0V). The source-drain voltage is biased at $V_{ds}$=100mV. The left inset shows the real-time resistance response of a graphene device ($V_g$=0V) to the exposure of ethanol. The right inset shows the typical scanning-electron-microscopy (SEM) images of a back-gated graphene device with a scale bar equal to 2μm. **b.** Normalized LFN spectra multiplied by frequency ($S_I/I^2 \times f$) in three SLG devices exposed to acetonitrile vapor. All three devices feature the same characteristic frequency $f_c$, showing excellent reproducibility of the noise response to chemical gases.

High-quality GNRs with a narrower width (e.g. sub-10nm) are expected to result in a larger separation among the noise peaks and larger noise amplitudes (for a smaller GNR area), which would make the noise peaks observable at 300K or higher temperatures. The gate-dependence of LFN can be employed to probe the change of GNR band structures, which can find its use in a broad range of sensing applications. For example, the LFN spectroscopy of GNR can detect the surface functionalization, the bio-molecule attachment and the strain variations, all of which alter the band structure of GNRs chemically, biologically or mechanically.

A recent study by Rumyanttsev *et al*. [75] has demonstrated the LFN-based selective gas sensing in back-gated graphene device. The room-temperature LFN spectra were collected in 1min after the device exposure to the chemical vapors with a well-controlled pressure (a degassing process was applied before switching the vapors). The data showed a discernible change of the LFN spectra due to the graphene exposure to chemical vapors (see Figure 9a). The noise spectra in open air were close to the 1/f shape; whereas most vapors introduced Lorentzian bulges over the 1/f noise background. The Lorentzian noise components can relate to the additional traps created by the gas molecules, which lead to the trapping/ detrapping processes with specific time constants [104-108]. Furthermore, the normalized LFN multiplied by frequency ($S_I/I^2 \times f$) featured a maximum at a characteristic frequency $f_c$ (see Figure 9b), which was different in different vapors (e.g. $f_c$~10-20Hz and 500-700Hz for tetrahydrofuran and acetonitrile, respectively). The LFN spectra were reproducible from multiple measurements, which can be used for reliable chemical sensing. The frequency $f_c$ of vapor-induced noise spectra, in combination with the real-time resistance changes (see the left inset of Figure 9a), can serve as distinctive signatures for highly-selective gas sensing by a single graphene device. This approach avoids the fabrication of a dense sensor array which requires specific functionalization for individual gases.

The idea of LFN-based metrology has also been implemented on the scanning-probe-microscopy (SPM) platform. Sung *et al*. [76] have recently developed a scanning noise microscopy (SNM) for graphene strip devices: a scanning Pt tip is contacted to the graphene surface to measure the current noise spectrum through it. The length-dependence of the LFN amplitude was analyzed using an empirical model, which extracted the noise contribution from the device channel. Importantly, this SPM method gave a two-dimensional noise mapping of the graphene strip device, which clearly indicated the spatial fluctuations of the graphene qualities (e.g. structural defects). The SNM method can be integrated in existing SPM technologies and achieve a spatial resolution as small as 1nm by optimizing the tip size. A high-resolution noise mapping can be useful in detecting the local surface qualities on graphene and many other materials, which can benefit fundamental research on nanoscale devices.

The development of variability-based sensing applications with graphene can bring research opportunities in multiple fields. For example, if a single-trap graphene system can be achieved, the statistical analysis of RTN can be employed to extract the trap information. For example, the spatial location of a single trap away from the graphene surface can be estimated by the gate dependence of the time constant ratios (i.e. $\tau_1/\tau_2$) [111]. And the RTN behavior under a magnetic field can detect the spin resonance of a single electron in graphene devices, which may provide interest for graphene spintronics [168]. The new role of graphene variabilities, employing the 'signal fluctuations' as the 'sensing signal', would attract both fundamental and practical interest.

## VI. CONCLUSION

This paper has reviewed the variability effects in graphene, with emphasis on their challenge and opportunities for device engineering and applications. The variabilities in as-made graphene devices can result from the environmental disturbance (Class I) and the material imperfection (Class II). In Class I variabilities, we review the research on interface traps near the graphene surface, with the focus on their physical principles, characterization methods and the approach to minimize their effect via device engineering. In Class II variabilities, we review the research on edge disorders that broadly exist in graphene materials, discussing their concepts, critical roles in device performance and the technology advancement in reducing their effect. From the metrology perspective, we discuss the potential use of graphene variabilities for sensing applications, such as the surface-quality detection, selective gas sensing and SPM technologies, which may promote interest in developing variability-based graphene applications.

Aligning the concepts of graphene variabilities with those in silicon devices, we see that the research is still at an early stage. Efforts need to be made in larger spatial scales with discussions from the circuit and system perspectives. Even within the device level, an exclusive coverage of this rapidly-growing field is difficult for its multidisciplinary nature. For example, structural defects (away from the edges) are important variabilities in bulk graphene (with a micron width), which are much concerned in graphene samples prepared by CVD-based transfer technology [92, 140]. Artificially generated structural defects (e.g. ion irradiation) have been found to lower the mobility, shift the Dirac point, increase the noise and modify the transport properties [69, 169-171]. However, the quality of CVD- graphene has been significantly improved in recent years [17, 23, 172]. The effects of structural defects on the performance of CVD-based graphene devices would need to be evaluated with consideration to the presence of other variabilities (e.g. charged impurities). In addition, large geometrical distortions in graphene, such as ripples, can be potentially implemented in strain and thermal engineering [85, 93]. A recent study reveals that a giant pseudo magnetic field (300 Tesla) can be achieved in graphene nanobubbles via strain engineering [173].

With the presence of variabilities, scalability of the graphene devices is a critical issue to evaluate their ultimate promise. Besides the resistance scaling as discussed before (Section III), scaling behaviors of the graphene devices, such as on/off ratio and transconductance, require continuous focus. For example, Sui *et al.* [66] have reported the role of disorder (e.g. charged impurities) to the size-scaling of minimum conductivity at Dirac point in single-layer graphene devices. Meric *et al.* [174] have characterized the length dependence of the high-bias transport in dual-gated graphene devices, which reveal the effect of interface traps to the output conductance and current saturation. How these graphene variabilities will be affected by size scaling is yet another important topic to investigate. Similar to silicon devices [99, 101], the scaling of graphene devices may increase the impact of graphene variabilities on device performance. As the device scales down, the variabilities with a small scale (e.g. interface traps, atomic-scale edge disorders) might become more influential than those with a large scale (e.g. ripples larger than 300nm can be less likely to exist in small devices [93]). The effect of graphene variabilities also depends on technology advances. For instance, the scattering rate due to charged impurities is lower in graphene devices on a boron nitride substrate than those on a Si/SiO$_2$ substrate [3]. Research on these aspects will help the exploration of the scaling limit in graphene electronics.

We finally mark that our discussions can extend to other thin-film, nanowire and nanotube devices, in all of which variabilities exist and need to be addressed for device applications. A controlled manipulation of these variabilities may lead to flexible metrology tools that can provide surprises in the future.

## ACKNOWLEDGMENT

The authors would like to thank C. M. Torres Jr., J. Bai, L. Liao, Y. Huang, M. Wang, E. B. Song, J. Tang, C. Zeng, S. Aloni, T. Kuykendall, F. Liu, F. Miao, X. Zhang, M. Y. Han, E. Rossi, S. Adam and E. H. Hwang for many stimulating discussions and technical supports.

# IEEE COPYRIGHT AND CONSENT FORM

To ensure uniformity of treatment among all contributors, other forms may not be substituted for this form, nor may any wording of the form be changed. This form is intended for original material submitted to the IEEE and must accompany any such material in order to be published by the IEEE.Please read the form carefully and keep a copy for your files.

TITLE OF PAPER/ARTICLE/REPORT, INCLUDING ALL CONTENT IN ANY FORM, FORMAT, OR MEDIA (hereinafter, "The Work"):**Variability Effects in Graphene: Challenges and Opportunities for Device Engineering and Applications**

COMPLETE LIST OF AUTHORS:**Xu, Guangyu; Zhang, Yuegang; Duan, Xiangfeng; Balandin, Alexander; Wang, Kang**

IEEE PUBLICATION TITLE (Journal, Magazine, Conference, Book):**Proceedings of the IEEE**

## COPYRIGHT TRANSFER

1. The undersigned hereby assigns to The Institute of Electrical and Electronics Engineers,Incorporated (the "IEEE") all rights under copyright that may exist in and to: (a) the above Work,including any revised or expanded derivative works submitted to the IEEE by the undersigned based on the Work; and (b) any associated written or multimedia components or other enhancements accompanying the Work.

## CONSENT AND RELEASE

2. ln the event the undersigned makes a presentation based upon the Work at a conference hosted or sponsored in whole or in part by the IEEE, the undersigned, in consideration for his/her participation in the conference, hereby grants the IEEE the unlimited, worldwide, irrevocable permission to use, distribute, publish, license, exhibit, record, digitize, broadcast, reproduce and archive, in any format or medium, whether now known or hereafter developed: (a) his/her presentation and comments at the conference; (b) any written materials or multimedia files used in connection with his/her presentation; and (c) any recorded interviews of him/her (collectively, the "Presentation"). The permission granted includes the transcription and reproduction ofthe Presentation for inclusion in products sold or distributed by IEEE and live or recorded broadcast ofthe Presentation during or after the conference.

3. In connection with the permission granted in Section 2, the undersigned hereby grants IEEE the unlimited, worldwide, irrevocable right to use his/her name, picture, likeness, voice and biographical information as part of the advertisement, distribution and sale ofproducts incorporating the Work or Presentation, and releases IEEE from any claim based on right of privacy or publicity.

4. The undersigned hereby warrants that the Work and Presentation (collectively, the "Materials") are original and that he/she is the author of the Materials. To the extent the Materials incorporate text passages, figures, data or other material from the works of others, the undersigned has obtained any necessary permissions. Where necessary, the undersigned has obtained all third party permissions and consents to grant the license above and has provided copies of such permissions and consents to IEEE.

[ ] Please check this box ifyou do not wish to have video/audio recordings made ofyour conference presentation.

See below for Retained Rights/Terms and Conditions, and Author Responsibilities.

## AUTHOR RESPONSIBILITIES

The IEEE distributes its technical publications throughout the world and wants to ensure that the material submitted to its publications is properly available to the readership of those publications. Authors must ensure that their Work meets the requirements as stated in section 8.2.1 of the IEEE PSPB Operations Manual, including provisions covering originality, authorship, author responsibilities and author misconduct. More information on

IEEEs publishing policies may be found at http://www.ieee.org/publications_standards/publications/rights/pub_tools_policies.html. Authors are advised especially of IEEE PSPB Operations Manual section 8.2.1.B12: "It is the responsibility of the authors, not the IEEE, to determine whether disclosure of their material requires the prior consent of other parties and, if so, to obtain it." Authors are also advised of IEEE PSPB Operations Manual section 8.1.1B: "Statements and opinions given in work published by the IEEE are the expression of the authors."

## RETAINED RIGHTS/TERMS AND CONDITIONS

### General

1. Authors/employers retain all proprietary rights in any process, procedure, or article of manufacture described in the Work.

2. Authors/employers may reproduce or authorize others to reproduce the Work, material extracted verbatim from the Work, or derivative works for the author's personal use or for company use, provided that the source and the IEEE copyright notice are indicated, the copies are not used in any way that implies IEEE endorsement of a product or service of any employer, and the copies themselves are not offered for sale.

3. In the case of a Work performed under a U.S. Government contract or grant, the IEEE recognizes that the U.S. Government has royalty-free permission to reproduce all or portions of the Work, and to authorize others to do so, for official U.S. Government purposes only, if the contract/grant so requires.

4. Although authors are permitted to re-use all or portions of the Work in other works, this does not include granting third-party requests for reprinting, republishing, or other types of re-use.The IEEE Intellectual Property Rights office must handle all such third-party requests.

5. Authors whose work was performed under a grant from a government funding agency are free to fulfill any deposit mandates from that funding agency.

### Author Online Use

6. Personal Servers. Authors and/or their employers shall have the right to post the accepted version of IEEE-copyrighted articles on their own personal servers or the servers of their institutions or employers without permission from IEEE, provided that the posted version includes a prominently displayed IEEE copyright notice and, when published, a full citation to the original IEEE publication, including a link to the article abstract in IEEE Xplore.Authors shall not post the final, published versions of their papers.

7. Classroom or Internal Training Use. An author is expressly permitted to post any portion of the accepted version of his/her own IEEE-copyrighted articles on the authors personal web site or the servers of the authors institution or company in connection with the authors teaching, training, or work responsibilities, provided that the appropriate copyright, credit, and reuse notices appear prominently with the posted material. Examples of permitted uses are lecture materials, course packs, e-reserves, conference presentations, or in-house training courses.

8. Electronic Preprints. Before submitting an article to an IEEE publication, authors frequently post their manuscripts to their own web site, their employers site, or to another server that invites constructive comment from colleagues.Upon submission of an article to IEEE, an author is required to transfer copyright in the article to IEEE, and the author must update any previously posted version of the article with a prominently displayed IEEE copyright notice. Upon publication of an article by the IEEE, the author must replace any previously posted electronic versions of the article with either (1) the full citation to the IEEE work with a Digital Object Identifier (DOI) or link to the article abstract in IEEE Xplore, or (2) the accepted version only (not the IEEE-published version), including the IEEE copyright notice and full citation, with a link to the final, published article in IEEE Xplore.

## INFORMATION FOR AUTHORS

### IEEE Copyright Ownership

It is the formal policy of the IEEE to own the copyrights to all copyrightable material in its technical publications and to the individual contributions contained therein, in order to protect the interests of the IEEE, its authors and their employers, and, at the same time, to facilitate the appropriate re-use of this material by others.The IEEE distributes its technical publications throughout the world and does so by various means such as hard copy, microfiche, microfilm, and electronic media.It also abstracts and may translate its publications, and articles contained therein, for inclusion in various compendiums, collective works, databases and similar publications.

**Author/Employer Rights**

If you are employed and prepared the Work on a subject within the scope of your employment, the copyright in the Work belongs to your employer as a work-for-hire. In that case, the IEEE assumes that when you sign this Form, you are authorized to do so by your employer and that your employer has consented to the transfer of copyright, to the representation and warranty of publication rights, and to all other terms and conditions of this Form. If such authorization and consent has not been given to you, an authorized representative of your employer should sign this Form as the Author.

## GENERAL TERMS

1. The undersigned represents that he/she has the power and authority to make and execute this form.
2. The undersigned agrees to identify and hold harmless the IEEE from any damage or expense that may arise in the event of a breach of any of the warranties set forth above.
3. In the event the above work is not accepted and published by the IEEE or is withdrawn by the author(s) before acceptance by the IEEE, the foregoing grant of rights shall become null and void and all materials embodying the Work submitted to the IEEE will be destroyed.
4. For jointly authored Works, all joint authors should sign, or one of the authors should sign as authorized agent for the others.

**Guangyu Xu**                                                                                                                     **15-02-2013**
**Author/Authorized Agent For Joint Authors**                                                     **Date(dd-mm-yy)**

THIS FORM MUST ACCOMPANY THE SUBMISSION OF THE AUTHOR'S MANUSCRIPT.
Questions about the submission of the form or manuscript must be sent to the publication's editor.Please direct all questions about IEEE copyright policy to:
IEEE Intellectual Property Rights Office, copyrights@ieee.org, +1-732-562-3966 (telephone)